\documentclass[prb,onecolumn,showpacs,preprintnumbers,amsmath,assume]{revtex4}
\usepackage{graphicx}
\usepackage{dcolumn}
\usepackage{bm}
\pdfoutput=1

\begin{document}

\title{Inversion of normal moveout for monoclinic media}

\author{Vladimir Grechka$^{1}$, Pedro Contreras$^{2}$, and Tsvankin Ilya$^{1}$}
\affiliation{$^{1}$ Center for Wave Phenomena, Department of Geophysics,Colorado School of Mines, Golden, CO 80401}
\affiliation{$^{2}$ PDVSA-INTEVEP, Apartado 76343, Caracas 1070A, Venezuela}
\date{\today}

\begin{abstract}
Multiple vertical fracture sets, possibly combined with horizontal fine
layering, produce an equivalent medium of monoclinic symmetry with
a horizontal symmetry plane.
Here, we show that multicomponent wide-azimuth reflection data (combined with known vertical velocity or reflector depth) or multi-azimuth walkaway VSP surveys provide enough information to invert for all but one anisotropic parameters of monoclinic media.

To facilitate the inversion procedure, we introduce a Thomsen-style parametrization
that includes the vertical velocities
of the $P$-wave and one of the split $S$-waves and a set of dimensionless
anisotropic coefficients. This notation
captures the combinations of the stiffnesses responsible for
the normal-moveout (NMO) ellipses of all three pure modes.
Our parameter-estimation algorithm, based on NMO equations valid for any
strength of the anisotropy, is designed to obtain anisotropic parameters
of monoclinic media by inverting the vertical velocities and NMO ellipses
of the waves $P$, $S_1$ and $S_2$.
A Dix-type representation of the NMO velocity of mode-converted waves makes
it possible to replace the pure shear modes in reflection surveys with
the waves $PS_1$ and $PS_2$. Numerical tests show that our method yields
stable estimates of all relevant parameters for both a single layer and
a horizontally stratified monoclinic medium.

\end{abstract}

\maketitle

\section{Introduction}

Natural fractures usually occur in vertical or subvertical sets (networks),
which makes fractured reservoirs azimuthally anisotropic with respect to
elastic wave propagation. In general, two or more sets of vertical
non-corrugated fractures produce an effective monoclinic medium with a horizontal
symmetry plane. Abundant geological evidence of multiple fracture sets corroborates potential importance of monoclinic models in seismic reservoir characterization.

Velocity analysis and parameter estimation for monoclinic media, however, is
a highly challenging task due to the large number of independent stiffness coefficients. To avoid ambiguity in the inversion procedure, we follow
the idea originally proposed for VTI media by \cite{Thomsen} and attempt to
identify the combinations of the stiffness coefficients responsible for
seismic signatures commonly measured from reflection data. Since moveout
velocity analysis is one of the most reliable tools for mapping the elastic
properties of the subsurface, we use the NMO ellipses of $P$-,
$S_1$, and $S_2$-waves \cite{Grechka1} from a horizontal reflector to define the anisotropic parameters of monoclinic media. Despite the absence of vertical symmetry planes in our
model, the polarization directions of the vertically propagating shear waves
establish a natural coordinate frame with the vanishing stiffness coefficient
$c_{45}$. We specify the anisotropic parameters in this coordinate system
(where the stiffness tensor has 12 independent elements) by analogy with the
Thomsen-style notation of \cite{Tsvankin} for orthorhombic media.
A subset of these parameters
($\epsilon^{\rm (1,2)}$, $\delta^{\rm (1,2)}$, and $\gamma^{\rm (1,2)}$) is
largely responsible for the values of the {\em semi-axes}$\,$ of the NMO ellipses.
Three newly introduced anisotropic coefficients $\zeta^{\rm (1,2,3)}$ mainly
describe the {\em rotation}$\,$ of the $P$- and $S$-wave NMO ellipses with
respect to the chosen coordinate frame.

We show that the vertical velocities and NMO ellipses of the $P$- and two split
$S$-waves can be used to estimate eleven parameters of monoclinic media.
The only parameter not constrained by conventional-spread moveout data from
horizontal reflectors is the anisotropic coefficient $\delta^{(3)}$.
We present numerical tests for a single layer and stratified monoclinic media
which confirm the accuracy of our inversion procedure and its stability with respect to errors in input data.

\section{Selection of coordinate frame}
Description of seismic wave propagation in monoclinic media with a
horizontal symmetry plane takes the simplest form in the coordinate
frame where the stiffness coefficient $c_{45}=0$. The $x_1$- and
$x_2$-axes of this coordinate system coincide with the polarization directions
of the vertically propagating $S$ waves.
Denoting the fast shear wave by $S_1$ and the slow one by $S_2$, we find the following expressions for
the vertical component $q$ of the slowness vector
${\bf p} = \{p_1, p_2, q\}$ for waves traveling in the $x_3$ (vertical) direction:

\begin{equation}
   q^{P} = \frac{1}{\sqrt{c_{33}}} \, ,
   q^{S_1} = \frac{1}{\sqrt{c_{55}}} \, ,
   q^{S_2} = \frac{1}{\sqrt{c_{44}}} \, .
\label{eq06}
\end{equation}
Here it is assumed that the $x_1$-axis points in the direction of the
polarization of the fast shear wave $S_1$, which implies that
$c_{55} > c_{44}$.

\section{NMO ellipses for horizontal reflectors}

Grechka and Tsvankin (1998) showed that  azimuthally varying NMO
velocity $V_{\rm nmo}(\alpha)$ of pure (non-converted) modes is represented by the following quadratic form that usually specifies an {\it ellipse}$\,$ in the horizontal plane:
\begin{equation}
   V_{\rm nmo}^{-2} (\alpha) =
              W_{11} \cos^2 \alpha +
         2 \, W_{12} \sin \alpha \cos \alpha +
              W_{22} \sin^2 \alpha \, .
\label{eq08}
\end{equation}
For a single horizontal layer of arbitrary symmetry, the matrix
${\bf W}$ is given by \cite{Grechka2}

\begin{equation}
  {\bf W} = { - \, {q} \over
              {q^{}_{,11} q^{}_{,22} - q_{,12}^2} } \,
        \left(
          \begin{array}{cc}
            q^{}_{,22} & - q^{}_{,12} \\
            - q^{}_{,12} & q^{}_{,11} \\
          \end{array}
        \right)
        \label{eq189}
\end{equation}
where $q \equiv q(p^{}_1, p^{}_2)$ denotes the vertical slowness
component,
$\left. q^{}_{,i} \equiv \frac{\partial q}{\partial p^{}_i} \right|_{p_i=0}$,
and
$\left. q^{}_{,ij} \equiv \frac{\partial^2 q}{{\partial p^{}_i} {\partial p^{}_j}}
\right|_{p_i= p_j=0}$.

Equivalently, the NMO ellipse~(\ref{eq08}) can be written through the eigenvalues $\lambda_{1,2}$ of the matrix ${\bf W}$ as
\cite{Grechka1}
\begin{equation}
   V_{\rm nmo}^{-2}(\alpha) = \lambda_1 \cos^2 (\alpha - \beta) +
                              \lambda_2 \sin^2 (\alpha - \beta) \, .
\label{eq091}
\end{equation}
Here
$\beta$ is the rotation angle of the ellipse with respect to the horizontal coordinate axes:
\begin{equation}
   \tan 2\beta = { {2 \, W_{12}} \over {W_{11} - W_{22}} } \, .
\label{eq092}
\end{equation}

The matrix {\bf W} for horizontal events in monoclinic media is obtained from
equation~(\ref{eq189}) by substituting the corresponding
values of $q$ equations~(\ref{eq06}) and the derivatives $q^{}_{,i}$ and
$q^{}_{,ij}$  which can be determined from the Christoffel equation. The exact
expressions for the matrices ${\bf W}^{P}$, ${\bf W}^{S_1}$, and
${\bf W}^{S_2}$ in a monoclinic layer in terms
of the stiffness coefficients $c_{ij}$ are rather lengthy. They do show,
however, that the ``purely monoclinic'' coefficients $c_{16}$, $c_{26}$, and $c_{36}$, which vanish in orthorhombic media, contribute to the diagonal elements
$W_{11}^{Q}$ and $W_{22}^{Q}$ ($Q = P$, $S_1$, or $S_2$) only through the
products $c_{i6} c_{j6}$ ($i,j=1,2,3$).

In contrast, the off-diagonal matrix elements $W_{12}^{Q}$ are approximately {\em linear}$\,$ in $c_{i6}$. Since the rotation angle
$\beta$ of the NMO ellipse is almost proportional to $W_{12}^{Q}$ [equation~(\ref{eq092})], the coefficients
$c_{i6}$ are primarily responsible for the rotation (but not for the semi-axes) of the NMO ellipses.

\section{Anisotropic parameters of monoclinic media and linearized NMO ellipses}

Below is a list of our Thomsen-type parameters for monoclinic media with a horizontal symmetry plane defined through the
density-normalized stiffness coefficients $c_{ij}$. Note that the vertical
velocities and anisotropic coefficients $\epsilon^{(i)}$, $\delta^{(i)}$ and
$\gamma^{(i)}$ are introduced exactly in the same way as the corresponding
\cite{Tsvankin} parameters for orthorhombic media with the vertical symmetry
planes $[x_1,x_3]$ and $[x_2,x_3]$.

\noindent $V_{P0}$ -- the $P$-wave vertical velocity:
      \begin{equation}
         V_{P0} \equiv  \sqrt{c_{33}} \, ;
      \label{eq11}
      \end{equation}
$V_{S0}$ -- the vertical velocity of the fast shear wave $S_1$ polarized in the $x_1$-direction:
      \begin{equation}
         V_{S0} \equiv  \sqrt{c_{55}} \, ;
      \label{eq12}
      \end{equation}
      \begin{equation}
         \epsilon^{(1)} \equiv { {c_{22} - c_{33}} \over
                                            {2 \, c_{33}} } \, ;
      \label{eq13}
      \end{equation}
      \begin{equation}
         \delta^{(1)} \equiv
         \frac{(c_{23}+c_{44})^2 - (c_{33} - c_{44})^2} {2 \, c_{33} \,
                       (c_{33}-c_{44})} \, ;
      \label{eq14}
      \end{equation}
      \begin{equation}
         \gamma^{(1)} \equiv { {c_{66} - c_{55}} \over
                        {2 \, c_{55}} } \, ;
      \label{eq15}
      \end{equation}
      \begin{equation}
         \epsilon^{(2)} \equiv { {c_{11} - c_{33}} \over
                    {2 \, c_{33}} } \, ;
      \label{eq16}
      \end{equation}
      \begin{equation}
         \delta^{(2)} \equiv
         \frac{(c_{13}+c_{55})^2 - (c_{33} - c_{55})^2} {2 \, c_{33} \,
              (c_{33}-c_{55})} \, ;
      \label{eq17}
      \end{equation}
      \begin{equation}
         \gamma^{(2)} \equiv { {c_{66} - c_{44}} \over
                   {2 \, c_{44}} } \, ;
      \label{eq18}
      \end{equation}
            \begin{equation}
         \delta^{(3)} \equiv
         \frac{(c_{12}+c_{66})^2 - (c_{11} - c_{66})^2}
              {2 \, c_{11} \, (c_{11}-c_{66})} \, ;
      \label{eq19}
      \end{equation}
      $\zeta^{(1)}$ -- the parameter responsible for the rotation of
      the $S_1$-wave NMO ellipse:
      \begin{equation}
         \zeta^{(1)} \equiv \frac{c_{16} - c_{36}}{2 \, c_{33}} \, ;
      \label{eq20}
      \end{equation}
      $\zeta^{(2)}$ -- the parameter responsible for the rotation of
      the $S_2$-wave NMO ellipse:
      \begin{equation}
         \zeta^{(2)} \equiv \frac{c_{26} - c_{36}}{2 \, c_{33}} \, ;
      \label{eq21}
      \end{equation}
      $\zeta^{(3)}$ -- the parameter responsible for the rotation of
      the $P$-wave NMO ellipse:
      \begin{equation}
         \zeta^{(3)} \equiv \frac{c_{36}}{c_{33}} \, .
      \label{eq22}
      \end{equation}
The anisotropic parameters $\zeta^{(1,2,3)}$ depend on the stiffness
elements that vanish in orthorhombic media. The coefficient $\zeta^{(3)}$ is
analogous to the anisotropic parameter $\chi_z$ introduced by \cite{Mensch}.
The parameters $\zeta^{(1)}$ and $\zeta^{(2)}$ are different from their
counterparts in \cite{Mensch} notation because the latter is
based on the approximate phase-velocity function.

\section{Approximations for the NMO ellipses}
To demonstrate that the anisotropic coefficients defined above are useful
in describing the NMO ellipses in monoclinic media, we linearize the exact
equation~(\ref{eq189}) in terms of $\zeta^{(1)}$, $\zeta^{(2)}$, and
$\zeta^{(3)}$. Also, additional linearizations in the parameters
$\epsilon^{(1,2)}$  $\delta^{(1,2)}$ and $\gamma^{(1,2)}$ were performed for
the elements $W_{12}^{Q}$ of all matrices ${\bf W}^{Q}$.
This yields the following approximate expressions for the matrices ${\bf W}$ in a monoclinic layer:
\begin{eqnarray}
\label{eq23a}
    W_{11}^{P} = \frac{1}{V_{P0}^2 \, (1 + 2 \, \delta^{(2)})} \,
    W_{12}^{P} =  - 2 \, \frac{\zeta^{(3)}}{V_{P0}^2} \, , \nonumber \\
    W_{22}^{P} =  \frac{1}{V_{P0}^2 \, (1 + 2 \, \delta^{(1)})} \, ;
\end{eqnarray}

\begin{eqnarray}
\label{eq23b}
     W_{11}^{S_1} = \frac{1}{V_{S_1}^2 \, (1 + 2 \, \sigma^{(2)})} \, ,
     W_{12}^{S_1} = - 2 \, \frac{\zeta^{(1)}}{V_{S_1}^2} \,
                      \left( \frac{V_{P0}}{V_{S_1}} \right)^2 \, , \nonumber \\
     W_{22}^{S_1} = \frac{1}{V_{S_1}^2 \, (1 + 2 \, \gamma^{(1)})} \, ;
\end{eqnarray}

\begin{eqnarray}
\label{eq23c}
    W_{11}^{S_2} = \frac{1}{V_{S_2}^2 \, (1 + 2 \, \gamma^{(2)})} \, ,
    W_{12}^{S_2} = - 2 \, \frac{\zeta^{(2)}}{V_{S_2}^2} \,
                      \left( \frac{V_{P0}}{V_{S_2}} \right)^2 \, , \nonumber \\
    W_{22}^{S_2} =  \frac{1}{V_{S_2}^2 \, (1 + 2 \, \sigma^{(1)})} \, .
\end{eqnarray}
Here
$\sigma^{(2)} \equiv (\frac{V_{P0}}{V_{S_1}})^2 (\epsilon^{(2)} - \delta^{(2)})$,
$\sigma^{(1)} \equiv  (\frac{V_{P0}}{V_{S_2}})^2 (\epsilon^{(1)} - \delta^{(1)})$,
$V_{S_1} = V_{S0}$, and
$V_{S_2} = V_{S0} \sqrt{ { {1 + 2 \, \gamma^{(1)}} \over
                           {1 + 2 \, \gamma^{(2)}} } } \,$;
$V_{S_1}$ and $V_{S_2}$ are the vertical velocities of the fast and slow shear waves, respectively.

The approximate diagonal elements $W_{11}^{Q}$ and
$W_{22}^{Q}$ in equations~(\ref{eq23a})--(\ref{eq23c}) are identical to the
corresponding {\it exact}$\,$ expressions for orthorhombic media given in \cite{Grechka3}.
The ``monoclinic'' coefficients $\zeta^{(1,2,3)}$ contribute to $W_{11}^{Q}$ and
$W_{22}^{Q}$ only through their products that were dropped during the linearization procedure.
In contrast, the off-diagonal matrix elements $W_{12}^{Q}$ in
equations~(\ref{eq23a})--(\ref{eq23c}) are linear in the coefficients
$\zeta^{(1,2,3)}$ and quadratic in the other anisotropic parameters.
In monoclinic media, all three $W_{12}^{Q} \ne 0$ and, in general, the NMO ellipses of the $P$-, $S_1$- and $S_2$-waves have different orientations.

\begin{figure}[ht]
\includegraphics[width = 2.5 in, height= 2.5 in]{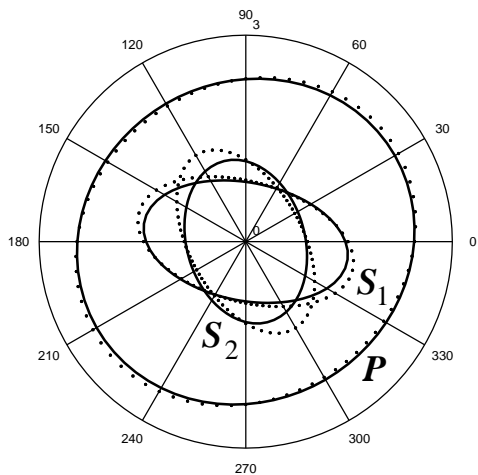}
\caption{Exact (solid) and approximate (dotted) NMO ellipses in a monoclinic layer with the following parameters:
         $V_{P0} = 2.0$ km/s,    $V_{S0} = 1.0$ km/s,
         $\epsilon^{(1)} = 0.3$, $\epsilon^{(2)} = 0.4$,
         $\delta^{(1)} = 0.2$,   $\delta^{(2)} = 0.25$,
         $\gamma^{(1)} = -0.1$,  $\gamma^{(2)} = 0.15$,
         $\zeta^{(1)} = -0.03$, $\zeta^{(2)} = -0.02$, $\zeta^{(3)} = 0.04$.
        }
\label{fig01}
\vspace{-0.3cm}
\end{figure}

Numerical tests show that the accuracy of the
approximations~(\ref{eq23a})--(\ref{eq23c}) depends mostly on the parameters
$\zeta^{(1,2,3)}$ and is less sensitive to the other anisotropic coefficients.
The example in Figure~\ref{fig01} demonstrates that for typical values of
$\zeta^{(1,2,3)}$ up to 0.05--0.06 equations~(\ref{eq23a})--(\ref{eq23c}) (dotted lines) yield a qualitatively
adequate approximation for the exact NMO ellipses (solid).

As predicted by equations~(\ref{eq23a})--(\ref{eq23c}), the axes of the NMO
ellipses of \mbox{$P$-,} $S_1$-, and $S_2$-waves in Figure~\ref{fig01} are
parallel neither to each other nor to the coordinate directions (i.e., to the $S$-wave polarizations).
This is a distinctive feature of monoclinic
media that can be used in the inversion for the anisotropic parameters.

\section{Parameter estimation}
\subsection{Analysis of the weak-anisotropy approximation}

Although equations~(\ref{eq23a})--(\ref{eq23c}) lose accuracy with increasing
$|\zeta^{(1,2,3)}|$, they provide a useful insight into the influence of the
medium parameters on normal moveout and help to design the inversion
procedure. Note that only one of the anisotropic coefficients, $\delta^{(3)}$,
is not contained in any of these equations and, therefore, cannot be estimated
from NMO velocities of horizontal events.
In the weak-anisotropy approximation,
the inversion of $W_{11}^{Q}$ and $W_{22}^{Q}$ is completely analogous to the
parameter-estimation problem in orthorhombic media. As discussed by \cite{Grechka3},
the vertical velocities, $W_{11}^{Q}$ and $W_{22}^{Q}$ can be inverted for the anisotropic coefficients $\epsilon^{(1,2)}$,
$\delta^{(1,2)}$, and $\delta^{(1,2)}$.
The remaining anisotropic parameters $\zeta^{(1,2,3)}$ of monoclinic media
can be estimated from the off-diagonal matrix elements $W_{12}^{Q}$, which
yield three
more equations for the three unknown parameters. We confirm these conclusions by performing numerical inversion based on the exact NMO equations.

\subsection{Numerical inversion}

\begin{figure}[ht]
\includegraphics[width = 2.5 in, height= 2.5 in]{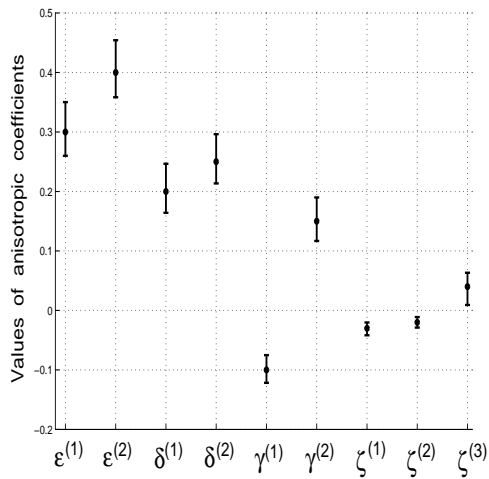}
\caption{Results of the moveout inversion for a monoclinic layer with the parameters specified in Figure~{\protect{\ref{fig01}}}.
The dots represent the exact values of the anisotropic parameters, the bars mark the $\pm$ standard deviation in each parameter and represent the 95\% confidence intervals. The standard deviations in the velocities $V_{P0}$ and $V_{S0}$ (not shown here) are 2.1\% and 2.0\%, respectively.}
\label{fig02}
\end{figure}

Input data for the parameter-estimation procedure include the vertical
velocities and NMO ellipses of the three pure modes ($P$, $S_1$, $S_2$)
determined from either wide-azimuth reflection data or walkaway VSP's.
The coordinate frame needed for moveout inversion can be established by performing Alford (1986) rotation of small-offset shear-wave data to identify the $S$-wave polarization directions at vertical incidence.
The anisotropic coefficients $\epsilon^{(1,2)}$, $\delta^{(1,2)}$, $\gamma^{(1,2)}$ and $\zeta^{(1,2,3)}$ are obtained by inverting the exact equation~(\ref{eq189}) for the matrix {\bf W}.
Figure~\ref{fig02} displays the inversion results for a monoclinic layer with
the parameters specified in Figure~\ref{fig01}.
The NMO velocities of
the three modes were computed for azimuths $0^\circ$, $45^\circ$, $90^\circ$,
and $135^\circ$ from equations~(\ref{eq08}) and~(\ref{eq189}).
To simulate errors in measured data, we added Gaussian errors with a variance of 2\% to the vertical and NMO velocities. Then we reconstructed the NMO ellipses for each realization of the data (distorted by noise) and carried out the inversion.

Overall, the stability of the inversion algorithm is quite satisfactory, but there is a substantial variation in the results from one anisotropic parameter to another. In particular, the error bars for $\zeta^{(1)}$ and $\zeta^{(2)}$ are much smaller
than those for $\zeta^{(3)}$ and the other anisotropic parameters.
This can be explained by the structure of equations~(\ref{eq23a})--(\ref{eq23c}) for $W_{12}^{S_1}$, $W_{12}^{S_2}$ and $W_{12}^{P}$. The expressions for $W_{12}^{S_k}$ ($k = 1,2$) have scaling factors $(V_{P0}/V_{S_k})^2$ that reach 4.0 and 6.5 for the model used in our numerical test. Therefore, the elements $W_{12}^{S_k}$ are much more sensitive to $\zeta^{(1)}$ and $\zeta^{(2)}$ than $W_{12}^{P}$ is to $\zeta^{(3)}$, which helps to recover $\zeta^{(1,2)}$ with a higher accuracy.

The inversion technique introduced above was also extended to horizontally
layered monoclinic media using the generalized Dix equation of \cite{Grechka2} that allows one to obtain the exact interval NMO
ellipses. Although both the Dix differentiation and polarization layer stripping of shear waves have known limitations with respect to the vertical resolution, they give stable results for coarse intervals with sufficient thickness.

\section{Conclusions}
Effective monoclinic media with a horizontal symmetry plane represent a general anisotropic model of hydrocarbon reservoirs with two or more vertical fracture systems. An analytic study of normal moveout in monoclinic media, presented here, leads to a Thomsen-style notation that captures the combinations of the stiffness coefficients responsible for the NMO ellipses of $P$- and $S$-waves. Natural horizontal coordinate directions for monoclinic models are associated with the orthogonal polarization vectors of the vertically traveling shear waves.

The anisotropic coefficients of monoclinic media can be separated into two distinctly different groups. The first group contains seven parameters ($\epsilon^{\rm (1,2)}$, $\delta^{\rm (1,2,3)}$, and $\gamma^{\rm (1,2)}$) defined identically to the corresponding \cite{Tsvankin} coefficients for orthorhombic media. While $\delta^{\rm (3)}$ has no influence on the NMO ellipses of $P$- and $S$-waves, the remaining six coefficients control the normal-moveout velocities in the coordinate directions $x_1$ and $x_2$.

Three additional anisotropic coefficients $\zeta^{\rm (1,2,3)}$ (the second parameter group) depend on the elements of the monoclinic stiffness tensor which vanish in orthorhombic media. The $\zeta$ coefficients determine the {\it rotation}$\,$ of the NMO ellipses with respect to the $S$-wave polarization directions.

Our algorithm, based on the exact NMO equations, is designed to invert the vertical velocities and NMO ellipses of the three pure modes (the shear waves can be replaced by mode conversions) for the anisotropic parameters $\epsilon^{\rm (1,2)}$, $\delta^{\rm (1,2)}$, $\gamma^{\rm (1,2)}$ and $\zeta^{\rm (1,2,3)}$.
Numerical tests for a single layer and stratified monoclinic media indicate that the inversion procedure is sufficiently stable, and all eleven parameters are well constrained by the vertical and NMO velocities.

\section{Acknowledgments}

We are grateful to members of the A(nisotropy)-Team of the Center for
Wave Phenomena (CWP) at CSM for helpful discussions. Pedro Contreras thanks
PDVSA-INTEVEP for giving him the opportunity to work at CWP.
V. Grechka and I. Tsvankin acknowledge the support provided by the members of the Consortium Project on Seismic Inverse Methods for Complex Structures at CWP and by the United States Department of Energy (award \#DE-FG03-98ER14908). I. Tsvankin was also supported by the Shell Faculty Career Initiation Grant.

\end{document}